\documentclass[useAMS,usenatbib]{mn2e}  
\usepackage {graphicx,subfig,aas_macros,float,amsmath}
\usepackage{ulem}
\usepackage[dvipsnames]{xcolor}  

\newcommand{\equ}[1]{eq.~(\ref{eq:#1})}
\newcommand{\equs}[1]{eqs.~(\ref{eq:#1})}
\newcommand{\equm}[1]{(\ref{eq:#1})}

\newcommand{\equnp}[1]{eq.~\ref{eq:#1}}

\newcommand{\se}[1]{\S\ref{sec:#1}}
\newcommand{\fig}[1]{Fig.~\ref{fig:#1}}
\newcommand{\figs}[1]{Figs.~\ref{fig:#1}}
\newcommand{\figss}[1]{\ref{fig:#1}}
\newcommand{\Fig}[1]{Figure~\ref{fig:#1}}

\newcommand{\be}{\begin{equation}}
\newcommand{\ee}{\end{equation}}
\newcommand{\bea}{\begin{eqnarray}}
\newcommand{\eea}{\end{eqnarray}}

\newcommand{\no}{\noindent}

\newcommand{\msun}{{\rm M}_\odot}

\newcommand{\ifm}[1]{\relax\ifmmode#1\else$\mathsurround=0pt #1$\fi}
\newcommand{\kms}{\ifmmode\,{\rm km}\,{\rm s}^{-1}\else km$\,$s$^{-1}$\fi}

\newcommand{\kpc}{\,{\rm kpc}}
\newcommand{\pc}{\,{\rm pc}}
\newcommand{\Gyr}{\,{\rm Gyr}}

\newcommand{\K}{\,{\rm K}}

\newcommand{\ltsima}{$\; \buildrel < \over \sim \;$}
\newcommand{\lsim}{\lower.5ex\hbox{\ltsima}}
\newcommand{\gtsima}{$\; \buildrel > \over \sim \;$}
\newcommand{\gsim}{\lower.5ex\hbox{\gtsima}}

\def\la{\langle}

\def\cmc{\,{\rm cm}^{-3}}
\def\cms{\,{\rm cm}^{-2}}

\def\M*{M_{\rm *}}
\def\Mv{M_{\rm v}}
\def\Rv{R_{\rm v}}
\def\Vv{V_{\rm v}}
\def\tv{t_{\rm v}}

\def\Tv{T_{\rm v}}
\def\Tb{T_{\rm b}}
\def\Ts{T_{\rm s}}
\def\Rs{R_{\rm s}}
\def\Rsv{{\cal R}_{\rm sv}}
\def\n0{n_{\rm H,0}}
\def\rhob{\rho_{\rm b}}
\def\rhos{\rho_{\rm s}}

\def\cb{c_{\rm b}}
\def\cs{c_{\rm s}}

\def\Vs{V_{\rm s}}

\def\tsc{t_{\rm sc}}
\def\tcm{t_{\rm cool,\,mix}}
\def\Mb{M_{\rm b}}
\def\Ms{M_{\rm s}}

\def\Pi{\varpi_{_{\rm I}}}

\usepackage{color}
\newcommand{\nmr}[1]{{#1}}

\begin{document} 

\large 

\title[Ly$\alpha$ Blobs from Cold Streams]
{Ly$\alpha$ Blobs from Cold Streams Undergoing Kelvin-Helmholtz Instabilities}

\author[Mandelker et al.] 
{\parbox[t]{\textwidth} 
{ 
Nir Mandelker$^{1,2}$\thanks{E-mail: nir.mandelker@yale.edu },
Frank C. van den Bosch$^1$,
Daisuke Nagai$^{1,3}$,
Avishai Dekel$^4$,
Yuval Birnboim$^4$,
Han Aung$^3$
} 
\\ \\ 
$^1$Department of Astronomy, Yale University, PO Box 208101, New Haven, CT, USA;\\
$^2$Heidelberger Institut f{\"u}r Theoretische Studien, Schloss-Wolfsbrunnenweg 35, 69118 Heidelberg, Germany;\\
$^3$Department of Physics, Yale University, New Haven, CT 06520, USA;\\
$^4$Centre for Astrophysics and Planetary Science, Racah Institute of Physics, The Hebrew University, Jerusalem 91904, Israel
}
\date{} 
 
\pagerange{\pageref{firstpage}--\pageref{lastpage}} \pubyear{0000} 
 
\maketitle 
 
\label{firstpage} 
 
\begin{abstract} 
\nmr{We present an analytic toy model for the radiation produced by the interaction between cold streams thought to feed massive halos at high redshift and their hot CGM. We begin by deriving cosmologically motivated parameters for the streams as they enter the halo virial radius, $\Rv$, as a function of halo mass and redshift. For $10^{12}\msun$ halos at $z=2$, we find the stream density to be $n_{\rm H,s}\sim (0.1-5)\times 10^{-2}\cmc$, a factor of $\delta \sim (30-300)$ times denser than the hot CGM, while stream radii are in the range $\Rs\sim (0.03-0.50)\Rv$. As streams accelerate towards the halo centre, they become denser and narrower. The stream-CGM interaction induces Kelvin-Helmholtz Instability (KHI), which leads to entrainment of CGM mass by the stream and to stream deceleration by momentum conservation. Assuming the entrainment rates derived by \citet{M20} in the absence of gravity can be applied locally at each halocentric radius, we derive equations of motion for the stream in the halo. Using these, we derive the net acceleration, mass growth, and energy dissipation induced by the stream-CGM interaction, as a function of halo mass and redshift, for different CGM density profiles. For the range of model parameters considered, we find that the interaction induces dissipation luminosities $L_{\rm diss}>10^{42}~{\rm erg~s^{-1}}$ within $\lsim 0.6\Rv$ of halos with $\Mv>10^{12}\msun$ at $z=2$. The emission scales with halo mass and redshift approximately as $\propto \Mv\,(1+z)^2$. The magnitude and spatial extent of the emission are consistent with observed Ly$\alpha$ blobs, though better treatment of the UV background and self-shielding is needed to solidify this conclusion.}
\end{abstract} 
 
 
\section{Introduction}
\label{sec:intro}

\smallskip
Hundreds of extended Ly$\alpha$ sources, known as Ly$\alpha$ blobs (LABs) have been observed at redshifts $z>2$ \citep{Steidel00,Steidel11,Palunas04,Matsuda04,Matsuda06,Matsuda11,Nilsson06,Saito06,Smith07,Yang09,Yang10}. These have luminosities of $L_{\rm Ly\alpha}\sim 10^{42}-10^{44}{\rm erg\,s^{-1}}$, though many fainter sources exist as well \citep{Rauch08,Steidel11}. Their spatial extents are typically several tens of kpc,  occasionally reaching up to $\sim 100\kpc$. Often there is no obvious central source capable of powering the Ly$\alpha$ emission, such as a starburst or AGN \citep{Saito06,Nilsson06,Yang09}, though in some cases there is evidence for an obscured central source \citep{Matsuda07,Geach07,Geach09,Prescott08}. LABs appear different from more recently detected giant Ly$\alpha$ nebulae \citep{Cantalupo14,Hennawi15,Borisova16,Martin16,Arrigoni18,Arrigoni19}. The latter are typically larger, extending $>100\kpc$, beyond the expected virial radius of their host dark matter halo, and associated with obvious AGN. 

\smallskip
The power source of observed LABs remains unclear. Some have speculated that all LABs are powered by a (potentially obscured) central AGN or starburst, either by photoionization or by ejection of superwinds into the halo  \citep{Haiman01,Ohyama03,Mori04,Weidinger04,Weidinger05,Wilman05,Laursen07,Geach09}. Another possibility is that cooling radiation of gas accreting onto the halo and/or onto the central galaxy fuels the LABs \citep{Fardal01,bd03,Furlanetto05,Nilsson06,Smith07,Dijkstra06,Dijkstra09,Goerdt10,FG10,Matsuda11}. 

\smallskip
The cooling radiation scenario is particularly intriguing in the context of the cold-stream model of galaxy formation, whereby massive galaxies at high redshifts are fed by narrow streams of dense gas which trace cosmic web filaments \citep{db06,Dekel09}. Owing to their high densities and short cooling times, the gas in these streams is not expected to shock at the virial radius. Rather, the streams maintain a temperature of $\Ts\gsim 10^4\K$ and are thought to penetrate the hot circumgalactic medium (CGM) and reach the central galaxy in roughly a halo crossing time. Such cold streams are ubiquitous in cosmological simulations \citep[e.g.][]{Keres05,Ocvirk08,Dekel09,CDB,FG11,vdv11}, where they are found to supply the halo with gas at rates comparable to the predicted cosmological accretion rate, with a significant fraction of the gas reaching the central galaxy \citep{Dekel09,Dekel13}. Many observed LABs and giant Ly$\alpha$ nebulae have filamentary morphologies, with spatial and kinematic properties consistent with predictions for cold streams \citep{Nilsson06,Saito06,Smith07,Matsuda11,Cantalupo14,Martin14a,Martin14b,Martin19,Borisova16,Fumagalli17,Leclercq17,Arrigoni18}. Absorption line studies of the CGM around high-redshift massive galaxies are also suggestive of dense, cold, inspiralling gas streams \citep{Fumagalli11,Goerdt12,vdv12,Bouche13,Bouche16,Prochaska14}.

\smallskip
\citet{Dijkstra09} presented an analytic toy model for Ly$\alpha$ radiation from cold streams. They found that under reasonable assumptions, motivated by cosmological simulations of the time, cooling radiation from cold streams could account for all the observed LABs. In their model, the Ly$\alpha$ emission is powered by the gravitational energy lost as the stream flows down the potential well of the dark matter halo. They found that if at least $\sim 20\%$ of this energy went into heating the stream and was subsequently radiated away, the resulting emission would resemble LABs. However, no clear mechanism for tapping into this energy was identified, and the model instead simply assumed that this could occur through a series of weak shocks. We emphasize, though, that no such shocks have been explicitly identified in the cosmological simulations. 

\smallskip
\citet{Goerdt10} presented a similar toy model for LABs resulting from gravitational heating of cold streams. Using AMR cosmological zoom-in simulations of $\sim 10^{12}\msun$ halos at $z\sim 2$ \citep{CDB}, they confirmed their model and found that cold streams resemble LABs in terms of luminosity, morphology, and extent. However, they also did not identify the mechanism by which the gravitational energy released by falling down the potential well was converted into radiation. 

\smallskip
\citet{FG10} analysed SPH cosmological zoom-in simulations, with comparable mass and redshift to those analysed by \citet{Goerdt10}. They found that the Ly$\alpha$ luminosity produced by cooling radiation in cold streams was one to two orders of magnitude lower than in luminous LABs. 
They argued that the main difference between their results and those of \citet{Goerdt10} was in the treatment of self-shielding of dense gas from the UV background. The simulations used in \citet{Goerdt10} assumed that gas with density $n>0.1\cmc$ was self shielded, though in their estimates of the Ly$\alpha$ emission they assumed that even lower density gas was in collisional ionization equilibrium. \citet{FG10}, based on radiative transfer calculations, assumed gas with $n>0.01\cmc$ to be self-shielded. This lowered the temperature of stream gas due to decreased UV heating, and thus lowered the overall Ly$\alpha$ luminosity. 

\smallskip
However, \citet{FG10} also acknowledged that it is plausible that differences in the hydrodynamic method, SPH vs AMR, contributed to the difference in the predicted emission, especially given the low resolution in the streams in both studies. The resolution in most state-of-the-art simulations is adaptive, such that the effective mass of each resolution element is fixed. The spatial resolution thus becomes very poor, typically $\sim \kpc$ scales, in the low density CGM near the virial radius \citep[e.g.][]{Nelson16}. 
\nmr{While this may be enough to resolve the largest and most diffuse streams (see \se{stream_prop} below), dense streams can be as narrow as a few kpc (\se{stream_prop}; \citealp{P18,M18}) and are therefore unresolved in cosmological simulations. Moreover, the different physical processes and complex subgrid models employed by different cosmological simulations make it difficult to gain a physical understanding of stream evolution and generalize results, which are found to be sensitive to the numerical approach. }
For example, early simulations using the moving mesh code \texttt{AREPO} \citep{Springel10,Vogelsberger12} found that streams heat-up and dissolve at $\gsim 0.5\Rv$ \citep{Nelson13}, while comparable Eulerian AMR \citep{CDB,Danovich15} and Lagrangian SPH \citep{Keres05,FG10} simulations found them to remain cold and collimated until $\sim 0.25\Rv$. 

\smallskip
To overcome these issues, several recent works have studied the evolution of cold streams using analytic models and idealized, high-resolution simulations. Initial work focusing on pure hydro instabilities found that sufficiently narrow streams would be disrupted by Kelvin-Helmholtz instabilities (KHI) within the CGM, prior to reaching the central galaxy \citep{M16,P18,M19}. However, subsequent work including either self-gravity \citep{Aung19}, magnetic fields \citep{Berlok19b}, or radiative cooling (\citealp{M20}; hereafter M20), found that each of these effects stabilizes the stream against disruption. 

\smallskip
Besides their importance for studying stream evolution and survival, the aforementioned studies identify a self-consistent dissipation mechanism acting on the stream, induced by its interaction with the hot CGM. In particular, M20, the only such study thus far to include radiative cooling, found that the formation of a turbulent mixing zone between the stream and the halo following the onset of KHI, resulted in halo gas cooling and condensing onto the stream. This led to the loss of kinetic energy from the stream and thermal energy from the background. M20 found that roughly half of this energy was radiated by gas with temperatures $T<5\times 10^4\K$, and would thus mostly be emitted as Ly$\alpha$. However, the results of M20 were based on analysis of an infinite stream,\footnote{In practice, a stream in a periodic box.} with constant density and cross section, and without external acceleration. None of these assumptions are expected to hold for realistic streams. Motivated by these results, we present in this paper a toy model generalising the results of M20 to account for the effect of a halo potential on the stream. Using this model, we make predictions for the total radiation produced by streams with cosmologically motivated properties within dark matter halos. 

\smallskip
The remainder of this paper is organized as follows. In \se{theory} we review the main conclusions of M20 regarding KHI in radiatively cooling streams and the associated energy dissipation rates. In \se{stream_prop} we derive cosmologically motivated scaling relations for stream properties as a function of halo mass and redshift. In \se{toy_model} we present our toy model for stream evolution in dark matter halos, and make predictions for the total amount of radiation that may result from the instability. In \se{caveat} we discuss limitations of our model, and we summarize our conclusions in \se{conc}.

\section{Theoretical Framework} 
\label{sec:theory} 

\smallskip
In this section, we summarize the main results of M20 regarding KHI in radiatively cooling streams, and the associated energy dissipation rates. The system we consider is a cylindrical stream with radius $\Rs$, density $\rhos$, and temperature $\Ts$, flowing with velocity $V_{\rm s}$ through a static background ($V_{\rm b}=0$) with density $\rhob$ and temperature $\Tb$. We define the Mach number of the flow with respect to the sound speed in the background, $\Mb=V_{\rm s}/\cb$, and the density contrast between the stream and the background, $\delta=\rhos/\rhob$. We assume that the stream and the background are in pressure equilibrium, and therefore $\Tb/\mu_{\rm b}=\delta\,\Ts/\mu_{\rm s}$, where $\mu_{\rm b}\sim \mu_{\rm s} \sim 0.6$ are the mean molecular weights in the background and stream respectively. For $\mu_{\rm s}=\mu_{\rm b}$, the Mach number with respect to the sound speed in the stream is $\Ms=V_{\rm s}/\cs=\delta^{1/2}\Mb$.

\smallskip
The shearing motion between the stream and the background induces KHI, which leads to the formation of a turbulent mixing region surrounding the stream\footnote{\nmr{On a fundamental level, the mixing seen in the numerical simulations referenced here is artificial, since the viscous scale is not resolved. The mixing between the two phases is likely driven by turbulence that, by construction, cascades down to the viscosity length which is the same as the diffusion length. The diffusion scale depends on the Coloumb interaction length of an electron (its mean free path before Thompson scattering). At $\rho\sim 10^{-26} {\rm gr~cm^{-3}}$ and $T\sim 10^5\K$, this is approximately $10^{-3}\pc$. The assumption is that in reality, the fluids will eventually mix at this scale, and that the dissipation rate (through cooling) is conserved throughout the inertial range, explaining why the simulations are converged in terms of the total dissipation rate (\citealp{Gronke18,Gronke20,Fielding20}; M20).}}. The typical density and temperature in this region are\footnote{\nmr{In practice, the mixing region contains a broad distribution of densities and temperatures rather than a charecteristic value (e.g. M20, figure 10). However, studies have shown that the evolution of the mixing region can be well described by considering a fluid with $\rho_{\rm mix}$ and $T_{\rm mix}$ as defined above (\citealp{Gronke18,Gronke20,Ji19}; M20), though other studies have explored different characteristic densities and temperatures \citep{Li_hopkins20,Hobbs20}. It has also been shown that when the mixing region is well resolved, it maintains pressure equilibrium with the unmixed phases even as it cools \citep{Fielding20}, consistent with the definitions of $\rho_{\rm mix}$ and $T_{\rm mix}$.}} \citep{Begelman90,Gronke18} 
\be 
\label{eq:nmix}
\rho_{\rm mix}\sim \left(\rhob \rhos\right)^{1/2} = \delta^{-1/2}\rhos,
\ee 
\be 
\label{eq:Tmix}
T_{\rm mix}\sim \left(\Tb \Ts \right)^{1/2} = \delta^{1/2}\Ts.
\ee
{\no}The cooling time in the mixing region is thus 
\be 
\label{eq:tcool_mix}
\tcm = \frac{k_{\rm B}T_{\rm mix}}{(\gamma-1)n_{\rm mix}\Lambda(T_{\rm mix})},
\ee
{\no}where $\gamma$ is the adiabatic index of the gas, $k_{\rm B}$ is Boltzmann's constant, $n_{\rm mix}$ is the particle number density in the mixing region, and $\Lambda(T_{\rm mix})$ is the cooling function evaluated at $T_{\rm mix}$.

\smallskip
In the non-radiative case \citep{P18,M19}, the turbulent mixing layer expands into both the stream and the background and its width is well approximated by 
\be 
\label{eq:non_rad_h_t}
h(t) = \alpha V_{\rm s} t,
\ee
{\no}where the dimensionless growth rate, $\alpha$, is \citep{Dimotakis91} 
\be 
\label{eq:alpha_non_rad}
\alpha \simeq 0.21\times \left[0.8{\rm exp}\left(-3 M_{\rm tot}^2\right)+0.2\right],
\ee
{\no}with $M_{\rm tot}=\Vs/(\cs+\cb)$. This approximation is an excellent fit for 2d planar slabs and also for 3d cylinders so long as $h\lsim \Rs$. We thus obtain the timescale for the mixing layer to grow to the size of the stream, i.e. $h=\Rs$, 
\be 
\label{eq:tshear_non_rad}
t_{\rm shear} = \frac{\Rs}{\alpha \Vs}.
\ee 
{\no}For $\delta\sim 100$ and $\Mb\sim 1$, we get $\alpha\sim 0.05$ and $\Vs\sim 10\cs$. In this case, $t_{\rm shear}$ is comparable to the stream sound crossing time, 
\be 
\label{eq:tsc}
\tsc = \frac{2\Rs}{\cs}.
\ee 

\smallskip
M20 found that a key parameter for determining stream evolution is the ratio $\tcm/t_{\rm shear}$\footnote{\nmr{There is some controversy in the literature over whether the relevant cooling time is $\tcm$ or $t_{\rm cool,\,hot}$, i.e. the cooling time in the hot phase \citep{Li_hopkins20}. We here adopt $\tcm$, following \citet{Gronke18,Gronke20}, and M20.}}. If $t_{\rm shear}<\tcm$, the evolution proceeds similarly to the non-radiative case studied by \citet{M19}, and the stream is eventually disrupted by KHI. However, if $\tcm<t_{\rm shear}$, then background gas entrained in the mixing layer cools and condenses onto the stream before it is disrupted by KHI. The stream thus remains cold, dense and collimated, and is not disrupted by hydrodynamic instabilities. Rather, the stream mass actually increases with time, as it entrains more and more gas from its hotter surroundings. Similar conclusions were reached in recent studies of high-velocity clouds in a hot CGM environment \citep{Gronke18,Gronke20}. The condition $\tcm=t_{\rm shear}$ leads to a critical stream radius, 
\be 
\label{eq:Rscrit}
R_{\rm s,crit}\simeq 0.3\kpc~ \alpha_{0.1}\,\delta_{100}^{3/2}\,\Mb\,\frac{T_{\rm s,4}}{n_{\rm s,0.01}\Lambda_{\rm mix,-22.5}},
\ee
{\no}where $T_{\rm s,4}=\Ts/10^4\K$, $n_{\rm s,0.01}=n_{\rm H,s}/0.01\cmc$, $\Lambda_{\rm mix,-22.5}=\Lambda(T_{\rm mix})/10^{-22.5}{\rm erg~s^{-1}~cm^3}$, $\delta_{100}=\delta/100$, and $\alpha_{0.1}=\alpha/0.1$. 

\smallskip
The ratio $\tcm/t_{\rm shear}=(\Rs/R_{\rm s,crit})^{-1}$, so streams with $\Rs>R_{\rm s,crit}$ grow in mass rather than dissolve. M20 derived an approximate expression for the entrainment rate of hot gas onto the stream. The cold mass-per-unit-length (hereafter line-mass) as a function of time is given by\footnote{See \citet{Gronke20} for a similar expression for the case of spherical cold clouds.}
\be 
\label{eq:mass_rad_approx}
m(t) = m_0\left(1+\dfrac{t}{t_{\rm ent}}\right),
\ee
{\no}where $m_0=\pi\Rs^2\rhos$ is the initial stream line-mass, and we have introduced the entrainment timescale, 
\be 
\label{eq:tent}
t_{\rm ent} = \frac{\delta}{2}\left(\frac{t_{\rm cool,s}}{\tsc}\right)^{1/4} \tsc, 
\ee 
{\no}where $t_{\rm cool,s}$ is the cooling time at the stream temperature and density. In practice, M20 assumed the stream to be in thermal equilibrium with a UV background, so the net cooling time at $\Ts$ is formally infinite. Therefore, $t_{\rm cool,s}$ is replaced by $t_{\rm cool,1.5\Ts}$, the cooling time at $T=1.5\Ts$, which is roughly the minimal cooling time. However, M20 note that any temperature in the range $\sim (1.2-2)\Ts$ works equally well. The density is given by assuming pressure equilibrium. 

\smallskip
As more mass is entrained in the flow, conservation of momentum causes the stream to decelerate. The velocity of the stream as a function of time is well approximated by\footnote{\nmr{When $\tcm<t_{\rm shear}$, the entrained background material very rapidly mixes with the stream material (M20, figures 3 and 12). In this case, the whole stream moves at roughly $V_{\rm s}$, save for a very narrow region near the outer edge of the mixing region where the velocity quickly drops to 0 towards the background. When $\tcm\gsim t_{\rm shear}$, the velocity distribution is much wider, and there can be a strong velocity gradient between the stream axis and its interface.}} 
\be 
\label{eq:decel_rad}
V_{\rm s}(t) = \frac{V_{\rm s,0}}{1+t/t_{\rm ent}}.
\ee
{\no}This reduces the kinetic energy per-unit-length associated with bulk laminar flow, which is well fit by 
\be 
\label{eq:Ek_rad}
E_{\rm k}(t) = \frac{E_{\rm k,0}}{1+t/t_{\rm ent}}.
\ee
{\no}In addition to the stream losing kinetic energy, the background gas entrained by the stream loses thermal energy, at a rate per-unit-length of
\be 
\label{eq:erad}
{\dot{E}}_{\rm th,b} = {\dot {m}}(e_{\rm s}-e_{\rm b}) \simeq -\frac{{\dot {m}}\cb^2}{\gamma(\gamma-1)} = -\frac{9 m_0 \cb^2}{10t_{\rm ent}},
\ee 
{\no}where $e_{\rm b}=P/[(\gamma-1)\rhob]$ is the thermal energy per unit mass of the background fluid which is larger than that in the cold component by a factor $\delta \gg 1$, $\gamma=5/3$ is the adiabatic index of the gas, and $\cb^2=\gamma P/\rhob$ is the adiabatic sound speed in the background. In the final equation we have used \equ{mass_rad_approx} to approximate ${\dot{m}}$. 

\smallskip
M20 found that the cooling radiation emitted per-unit-length by the stream is very well approximated by 
\be 
\label{eq:Ldiss}
\mathcal{L}_{\rm diss}\simeq \left|\dot{E}_{\rm k}\right| + \frac{5}{3}\left|\dot{E}_{\rm th}\right|.
\ee
{\no}The factor $5/3$ accompanying $\dot{E}_{\rm th}$ accounts for the fact that the pressure of the background gas just outside the mixing layer remains roughly constant, and therefore the condensation of background gas onto the stream is well approximated as an isobaric cooling flow. In this case, the emitted radiation is given by the difference in enthalpy, rather than energy, between the initial and final states \citep{Fabian94}. In \equ{Ldiss} we have ignored any net heating of the stream, which is found to be very small (M20). For stream temperatures of order $\Ts\sim 10^4\K$ and background temperatures of order $\Tb\sim 10^6\K$, roughly half the luminosity is emitted at temperatures $T<5\times 10^4\K$, and is thus expected to contribute substantially to Ly$\alpha$ radiation (M20).

\section{Cosmologically Motivated Stream Properties}
\label{sec:stream_prop} 

\smallskip
In \citet{P18}, section 5.1, we estimated $\Mb$, $\delta$, and $\Rs/\Rv$ for streams near the virial radius of a halo supporting a virial shock. We revisit this calculation here, both to correct an error in the previous estimate of $\Rs/\Rv$, and because in the context of radiatively cooling streams we require additional parameters, such as the stream density, which were not addressed previously. For an alternative derivation of stream properties in halos without a virial shock, see \citet{M18}.

\smallskip
The stream velocity as it enters the halo is proportional to the virial velocity, $V_0=\eta \Vv$, with $0.5\lsim \eta \lsim 1$ \citep{Goerdt15a}. The maximal value $\eta$ is likely to obtain is $\sqrt{2}$ if the stream enters the halo at the escape velocity. The virial velocity and radius at $z\gsim 1$ are given by \citep{Dekel13}
\be 
\label{eq:Vvir}
\Vv\simeq 200\kms~M_{12}^{1/3}\,(1+z)_3^{1/2},
\ee
\be 
\label{eq:Rvir}
\Rv\simeq 100\kpc~M_{12}^{1/3}\,(1+z)_3^{-1},
\ee
{\no}with $M_{12}=\Mv/10^{12}\msun$ and $(1+z)_3=(1+z)/3$. The halo temperature at $\Rv$ is proportional to the virial temperature, $T_{\rm h}=\Theta_h \Tv$, with $\Theta_{\rm h}\gsim 3/8$ given by the jump conditions at the virial shock \citep{db06}, though it may be larger due to additional heating by feedback from the central or satellite galaxies, or when the virial shock propagates outwards. The virial temperature is given by
\be 
\label{eq:Tvir}
\Tv\simeq 1.5\times 10^6\K~M_{12}^{2/3}\,(1+z)_3.
\ee
{\no}The sound speed in the halo is 
\be 
\label{eq:cvir}
c_{\rm h} = \sqrt{\frac{\gamma k_{\rm B}T_{\rm h}}{\mu_{\rm h} m_{\rm p}}}\simeq 185\kms M_{12}^{1/3}\, (1+z)_3^{1/2} \,\Theta_{\rm h}^{1/2},
\ee
{\no}where we have assumed that $\mu_{\rm h}\sim 0.6$, appropriate for fully ionized gas with roughly primordial composition. The stream Mach number upon entering the halo is thus 
\be 
\label{eq:Mb_isothermal}
\Mb=V_0/c_{\rm h}\simeq 1.1~\eta\,\Theta_{\rm h}^{-1/2}.
\ee
{\no}Taking uncertainties in $\eta$ and $\Theta_{\rm h}$ into account, we have that $0.75\lsim \Mb\lsim 2.25$. 

\smallskip
If the stream is in thermal equilibrium with the UV background at $z\sim 2$, its temperature is $\Ts\sim 1.5\times 10^4$ (M20), with a mild dependence on density and \nmr{metallicity}. In practice, the temperature can be lower if the stream is self-shielded, or higher if the stream is highly turbulent before entering the halo, with turbulent dissipation timescales comparable to the cooling time. We therefore assume $\Ts=\Theta_{\rm s}1.5\times 10^4\K$, with $\Theta_{\rm s}\sim 0.5-2$\footnote{\nmr{We note that much lower values of $\Theta_{\rm s}$ could be possible if the stream were dense enough and/or metal-rich enough to allow cooling to much lower temperatures in less than a halo crossing time. While this may be the case at $z>5$, it is unlikely to occur at redshifts $z<4$ unless the \nmr{metallicity} in the stream is $\sim 0.1Z_{\odot}$ \citep{M18}. This is very high compared to values found in cosmological simulations and observed in both LABs and Lyman limit systems at these redshifts.}}. The sound speed in the stream is 
\be 
\label{eq:cs_model}
c_{\rm s} = \sqrt{\frac{\gamma k_{\rm B}T_{\rm s}}{\mu_{\rm s} m_{\rm p}}}\simeq 18.5\kms \,\Theta_{\rm s}^{1/2}\,\mu_{\rm s,0.6}^{-1/2},
\ee
{\no}with $\mu_{\rm s,0.6}=\mu_{\rm s}/0.6$. If the stream is strongly self-shielded and predominantly neutral, then $\mu_{\rm s,0.6}\sim 2$. Assuming the stream and the hot halo are in pressure equilibrium, the density contrast is given by 
\be 
\label{eq:delta_model}
\delta=\rhos/\rho_{\rm h}=T_{\rm h}/\Ts\simeq 100~M_{12}^{2/3}\,(1+z)_3\left(\Theta_{\rm h}/\Theta_{\rm s}\right).
\ee
{\no}Taking into account uncertainties in $\Theta_{\rm h}$ and $\Theta_{\rm s}$, density contrasts of $\delta\sim (30-300)$ are reasonable for a $\Mv\sim 10^{12}\msun$ halo at $z\sim 2$. 

\begin{figure*}
\begin{center}
\includegraphics[trim={0.02cm 0.03cm 0.01cm 0.01cm}, clip, width =0.99 \textwidth]{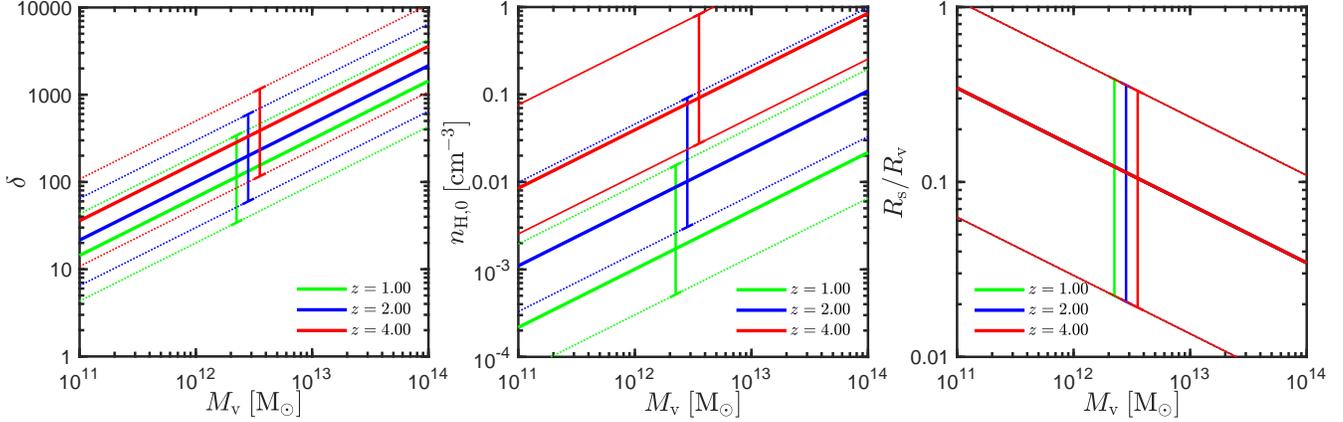}
\end{center}
\caption{Stream properties upon entering the halo virial radius, according to our model. We show, as a function of halo mass on the $x$ axis, the density contrast between the stream and the background, $\delta$ (\equnp{delta_model}, left), the stream Hydrogen number density, $n_{\rm H,0}$ (\equnp{n0}, centre), and the ratio of stream radius to halo virial radius, $\Rs/\Rv$ (\equnp{Rs_Rv_model}, right). Green, blue, and red lines represent redshifts $z=1$, $2$, and $4$ respectively. The thin lines spanned by error bars represent the range of stream properties obtained by varying the model parameters within the ranges $\Theta_{\rm h}\in(3/8,1.0)$, $\Theta_{\rm s}\in(0.5,2.0)$, $\eta\in(0.5,\sqrt{2})$, $\widetilde{f}_{\rm h}\in (1.0,3.0)$, and $\widetilde{s}\in(0.3-3.0)$ (see text for details). The thick lines represent our fiducial model, where all the above parameters have values of $1.0$. Note that $\Rs/\Rv$ has no redshift dependence in our model. }
\label{fig:params} 
\end{figure*}

\smallskip
The stream density upon entering the halo virial radius, $\rho_0$, is given by the density contrast, $\delta$, and the density of the hot halo gas at $\Rv$. The latter is given by 
\be 
\label{eq:rhob}
\rho_{\rm h}(\Rv) = \Delta(\Rv){\overline{\rho}_{\rm h}}\simeq \Delta(\Rv)f_{\rm h}f_{\rm b}18\pi^2 \overline{\rho}_{\rm u}(z),
\ee
{\no}where ${\overline{\rho}_{\rm h}}$ is the mean density of the hot component in the halo, ${\overline{\rho}_{\rm u}}(z)$ is the mean density of the universe at redshift $z$, $f_{\rm b}\simeq 0.17$ is the universal baryon fraction, and $f_{\rm h}$ denotes the fraction of the baryonic mass within the halo which is in the hot gas component. Simulations suggest this is $0.3\lsim f_{\rm h}\lsim 0.4$ for halo masses $M_{12}\sim (0.5-2)$ at redshift $z\gsim 2$ \citep{Roca19}, while observations of $M_{12}\sim (1-10)$ halos at $z\gsim 0.1$ suggest $f_{\rm h}\sim 0.2-0.3$ \citep{Singh18}. We define $f_{\rm h,0.3}=f_{\rm h}/0.3$. $\Delta(\Rv)$ is the ratio of the density at $\Rv$ to the mean density in the halo, ${\overline{\rho}_{\rm v}}=3\Mv/(4\pi\Rv^3)$. For a singular isothermal sphere $\Delta(\Rv)=1/3$, while for an NFW halo with concentration $c=(5,10,20)$, $\Delta(\Rv)\sim (0.24, 0.19, 0.14)$ respectively. We write $\Delta_{1/6}=\Delta(\Rv)/(1/6)$. Assuming cosmological parameters $\Omega_{\rm m}=0.3$ and $H_0=70\,{\rm km/s/Mpc}$, we obtain for the stream density at $\Rv$
\be 
\label{eq:rho0}
\rho_0 \simeq 1.1\times 10^{-26}{\rm gr\,cm^{-3}}\,(1+z)_3^3\,\delta_{100}\,{\widetilde{f}}_{\rm h},
\ee
{\no}where $\delta_{100}=\delta/100$, and ${\widetilde{f}}_{\rm h}=\Delta_{1/6} f_{\rm h,0.3}\sim (1-3)$. For a Hydrogen mass fraction of $X=0.76$ this yields a Hydrogen number density in the stream at $\Rv$ of 
\be 
\label{eq:n0}
\n0 \simeq 5.1\times 10^{-3}\cmc\,(1+z)_3^3\,\delta_{100}\,\widetilde{f}_{\rm h}.
\ee
{\no}We hereafter write $n_{\rm s,0.01}=\n0/(0.01\cmc)$.

\smallskip
The stream radius at $\Rv$ can be constrained using the mass flux through the stream, 
\be 
\label{eq:mdots}
{\dot {M}}_{\rm s}=\pi \Rs^2 \rho_0 V_0 = f_{\rm s}f_{\rm b}{\dot {M}}_{\rm v},
\ee 
{\no}where ${\dot {M}}_{\rm v}$ is the total mass accretion rate onto the halo virial radius, and $f_{\rm s}$ is the fraction of baryonic accretion along the gas stream. Cosmological simulations suggest $f_{\rm s}\sim (0.2-0.5)$ with a typical value of $f_{\rm s}=1/3$ \citep[i.e. three significant streams;][]{Danovich12}. In the Einstein de Sitter (EdS) regime (valid at $z>1$), the accretion onto the virial radius is well approximated by \citet{Dekel13} 
\be 
\label{eq:mdotv}
{\dot {M}}_{\rm v}/\Mv\simeq 0.47\Gyr^{-1}\,s\,(1+z)_3^{5/2},
\ee 
{\no}with halo-to-halo variance encapsulated by the normalization $s\sim (0.5-2)$. Combining \equs{Vvir}, \equm{rho0}, \equm{mdots}, and \equm{mdotv}, we obtain 
\be 
\label{eq:Rs_model}
\Rs \simeq 16\kpc~M_{12}^{1/3}\,(1+z)_3^{-1/2}\,\delta_{100}^{-1/2}\,\left(\frac{\widetilde{s}}{\eta \widetilde{f}_{\rm h}}\right)^{1/2},
\ee 
{\no}with $\widetilde{s}=sf_{\rm s}/(1/3)\sim(0.3-3)$. Together with \equ{Rvir} we have
\be 
\label{eq:Rs_Rv_model}
\frac{\Rs}{\Rv} \simeq 0.16~(1+z)_3^{1/2}\,\delta_{100}^{-1/2}\,\left(\frac{\widetilde{s}}{\eta \widetilde{f}_{\rm h}}\right)^{1/2}.
\ee
{\no}Note that this is larger by a factor $\sim 2$ than the corresponding equation (68) in \citet{P18}, which seems to be due to a typo as it is consistent with their equation (67). Using \equ{delta_model} for $\delta$, we see that $\Rs/\Rv\propto \Mv^{-1/3}$ and is independent of redshift. More massive halos are thus fed by relatively narrower streams compared to their virial radii. We hereafter substitute $\Rsv\equiv \Rs/\Rv$.

\smallskip
The Hydrogen column density perpendicular to the stream axis is $N_{\rm H,0}=\n0\Rs$. Using \equs{n0} and \equm{Rs_model} this yields
\be 
\label{eq:NH_model}
N_{\rm H,0}\simeq 3\times 10^{20}\cms M_{12}^{1/3}\,(1+z)_3^{5/2}\,\delta_{100}^{1/2}\left(\frac{\widetilde{s}\widetilde{f}_{\rm h}}{\eta}\right)^{1/2}.
\ee
{\no}For $10^{12}\msun$ halos at $z\sim 2$, this is large enough for the streams to be largely self-shielded against the UV background if the neutral fraction is $x_{\rm HI}\gsim 10^{-3}$. In collisional ionization equilibrium (CIE), this is true for $\Ts\lsim 3\times 10^4\K$ \citep[e.g.][]{Goerdt10}, i.e. $\Theta_{\rm s}\lsim 2$.
Of course, streams may still be susceptible to local sources of UV radiation, such as star-formation in satellite galaxies located along the stream, or starbursts/AGN activity in the central galaxy.

\smallskip
To summarize, given the halo mass and redshift, the precise temperatures of the stream and halo gas, $\Theta_{\rm s}$ and $\Theta_{\rm h}$ respectively, determine the density contrast, $\delta$. The stream Mach number, $\Mb$ is then interchangeable with $\eta$, the ratio of stream to virial velocity. The stream density is then determined by $\widetilde{f}_{\rm h}$, the normalization of the hot halo gas density near $\Rv$. Finally, the ratio of stream radius to virial radius is set by $\widetilde{s}$, the normalization of the gas accretion rate along the stream. For $\eta\sim 1$, $\widetilde{s}\sim(0.3-3)$ and ${\widetilde{f}}_{\rm h}\sim (1-3)$, a stream at $z=2$ has $\Rsv\sim (0.09-0.50)$, $(0.05-0.28)$, and $(0.03-0.16)$ for $\delta=30$, $100$, and $300$ respectively. The corresponding stream densities are $n_{\rm s,0.01} \sim (0.15-0.45)$, $(0.5-1.5)$, and $(1.5-4.5)$, with larger values of $\n0$ corresponding to smaller values of $\Rsv$. For a given $\delta$, neither property depends on halo mass. 

\smallskip
\Fig{params} shows the full plausible range of $\delta$, $\n0$, and $\Rsv$ as a function of halo mass at redshifts $z=1$, $2$, and $4$. As motivated above, we allow the model parameters to vary in the range $\Theta_{\rm h}\in(3/8,1.0)$, $\Theta_{\rm s}\in(0.5,2.0)$, $\eta\in(0.5,\sqrt{2})$, $\widetilde{f}_{\rm h}\in (1.0,3.0)$, and $\widetilde{s}\in(0.3-3.0)$, while in our fiducial model all these parameters have values of $1.0$. As noted above, inserting \equ{delta_model} into \equ{Rs_Rv_model} results in $\Rsv$ depending only on halo mass and not on redshift, as seen in the right-hand panel. In \fig{params}, we extend our model down to halo masses of $\Mv=10^{11}\msun$. While such halos are below the critical mass for forming a stable accretion shock at $\Rv$ \citep{bd03}, high-resolution simulations suggest that stellar feedback from the central galaxy can result in a quasi-stable hot CGM in halos of this mass as well \citep{Fielding17}. In lower mass halos, a hot CGM is not expected invalidating the fundamental assumption of our model, pressure equilibrium between cold streams and a quasi-stable hot atmosphere. At the high-mass end, we show results up to $\Mv=10^{14}\msun$. Streams penetrating such massive halos may be pre-heated in the IGM prior to entering the halo \citep{db06,Birnboim16}, particularly at $z\sim 1$. This would result in very large values of $\Theta_{\rm s}$ not considered here, so these results shoule be treated with caution.


\section{Stream Evolution and Energy Dissipation in Dark Matter Halos}
\label{sec:toy_model}

Here we outline our toy model for the evolution of streams in dark matter halos as they make their way towards the central galaxy. This is an extension of similar models presented in \citet{P18} (Appendix F) and \citet{M19} (section 5.3), now accounting for radiative cooling and energy dissipation, as well as self-consistently accounting for mass growth and deceleration of the stream in the equations of motion. We begin in \se{model_survive} by addressing stream survival within dark matter halos. In \se{model_scaling} we discuss how stream properties scale with halocentric radius. In \se{model_motion} we derive equations of motion for the stream as it penetrates the halo, and in \se{model_emis} we evaluate the dissipation and resulting luminosity induced by the stream-halo interaction. In \se{model_sol} and \fig{toy_model} we present example solutions of our model.

\begin{figure}
\begin{center}
\includegraphics[trim={0.01cm 0.03cm 0.01cm 0.01cm}, clip, width =0.48 \textwidth]{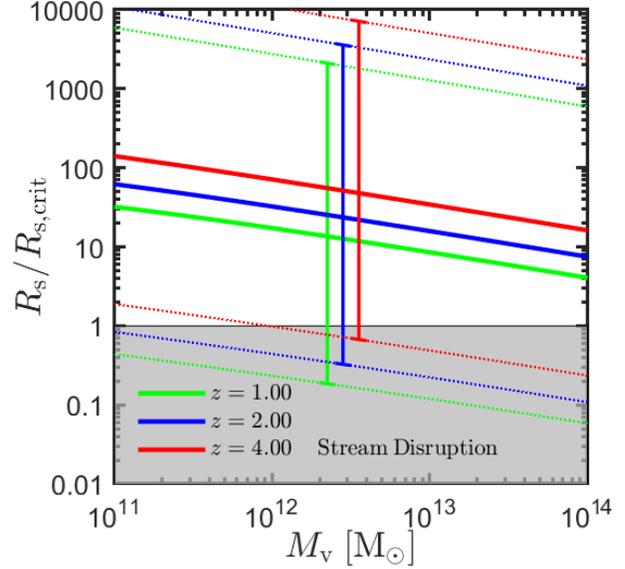}
\end{center}
\caption{As in \fig{params}, but showing the ratio of stream radius to the critical radius for cooling to dominate over KHI, $\Rs/R_{\rm s,crit}$ (\equnp{Rs_Rscrit}). Streams with $\Rs>R_{\rm s,crit}$ are likely to survive their journey through the CGM towards the central galaxy while remaining cold and coherent, while those with $\Rs<R_{\rm s,crit}$ (grey shaded region) are susceptible to disruption by KHI. We assume the same parameter ranges as in \fig{params}, and evaluate $\alpha_{0.1}$ for each parameter combination using \equ{alpha_non_rad}. We further assume $\Lambda_{\rm mix,-22.5}\in(0.5,2.0)$, with a fiducial value of 1.0. While $\Rs/R_{\rm s,crit}$ spans almost 4 orders of magnitude at each halo mass and redshift, there is a trend for the ratio to decrease towards larger halo masses and lower redshifts, making such streams more susceptible to disruption. }
\label{fig:Rscrit} 
\end{figure}

\subsection{Stream Survival in Halos}
\label{sec:model_survive}

\smallskip
We begin by addressing the survival of cold streams in hot halos, which is necessary for the validity of our model. This issue was addressed by M20, who found that cold streams are likely to survive the journey from the virial radius to the central galaxy, but can be strengthened here using our more accurate estimate of $\Rs$ (\equnp{Rs_model}). Based on the results of M20, summarized in \se{theory}, streams with $\Rs<R_{\rm s,crit}$ (\equnp{Rscrit}), i.e. where $\tcm>t_{\rm shear}$, will be disrupted by KHI similar to the non-radiative case described in \citet{M19}, while streams with $\Rs>R_{\rm s,crit}$ will survive. Inserting \equ{n0} and $T_{\rm s,4}=1.5\Theta_{\rm s}$ into \equ{Rscrit}, 
\be 
\label{eq:Rscrit_model}
R_{\rm s,crit} \simeq 0.9\kpc~(1+z)_3^{-3}\,\delta_{100}^{1/2}\,\frac{\alpha_{0.1}\,\Mb\,\Theta_{\rm s}}{\widetilde{f}_{\rm h}\,\Lambda_{\rm mix,-22.5}}.
\ee
{\no}Together with \equ{Rs_model}, we have 
\be 
\begin{array}{c}
\label{eq:Rs_Rscrit}
\dfrac{\Rs}{R_{\rm s,crit}} \simeq 18~M_{12}^{1/3}\,(1+z)_3^{5/2}\,\delta_{100}^{-1} \\
\\
\times \, \dfrac{\widetilde{s}^{1/2}\,\widetilde{f}_{\rm h}^{1/2}\,\Lambda_{\rm mix,-22.5}}{\eta^{1/2}\,\alpha_{0.1}\,\Mb\,\Theta_{\rm s}}.
\end{array}
\ee
{\no}This is $\gsim 1$, even in the extreme case where the \nmr{metallicity} in the mixing layer is 0, so $\Lambda_{\rm mix,-22.5}\sim 0.4$ for $T_{\rm mix}\sim 1.5\times 10^5\K$, $n_{\rm mix}\sim 5\times 10^{-4}\cmc$ and a $z=2$ \citet{HaardtMadau96} UV background. In this case, for $\delta=300$, $\widetilde{s}=0.3$, and $\widetilde{f}_{h}=1$, we obtain $\Rs/R_{\rm s,crit}\sim 1.3$ for $M_{12}=(1+z)_3=\Mb=\eta=\Theta_{\rm s}=\alpha_{0.1}=1$. This suggests that for halos with $\Mv\gsim 10^{12}\msun$ at $z\sim 2$, $\Rs>R_{\rm s,crit}$ in virtually all cases.

\smallskip
Inserting \equ{delta_model} into \equ{Rs_Rscrit}, we find that $\Rs/R_{\rm s,crit}\propto M_{12}^{-1/3}(1+z)_3^{3/2}$. The ratio thus decreases in more massive halos at lower redshift, making streams in such halos more susceptible to disruption. This is \nmr{qualitatively} consistent with the notion that cold streams do not penetrate very massive haloes at $z\lsim 2$ \citep[e.g.][]{db06,vdv11}. In \fig{Rscrit} we show the ratio $\Rs/R_{\rm s,crit}$ as a function of halo mass for $z=1$, $2$, and $4$, for the same range of parameters considered in \fig{params}. For each set of values for $\Mb$ and $\delta$, we used \equ{alpha_non_rad} to evaluate $\alpha_{0.1}$. We considered $\Lambda_{\rm mix,-22.5}\in(0.5,2.0)$, with a fiducial value of 1.0, though evaluating the cooling rate in the mixing layer for each set of parameters does not qualitatively change the picture for \nmr{metallicity} values $\lsim 0.1Z_{\odot}$. While the fiducial models always yield $\Rs/R_{\rm s,crit}>1$, there is a small range of parameters which yield $\Rs/R_{\rm s,crit}<1$, particularly at $z=1$. \nmr{Furthermore, for $\Rs/R_{\rm s,crit}<10$ the dominance of cooling over KHI becomes somewhat marginal (M20) and streams may heat up. This occurs for the fiducial model at $z=1$ for $\Mv\gsim 5\times 10^{12}\msun$.}

\smallskip
The above discussion is valid as the stream enters the halo at $\Rv$, where \equs{n0} and \equm{Rs_model} for the stream density and radius are applicable. As discussed in \se{model_scaling} below, the stream density increases at smaller halocentric radii. Ignoring variations in the stream temperature and velocity within the halo, which are expected to be small as justified below, we have $\Rs\propto n_{\rm s}^{-1/2}$ while $R_{\rm s,crit}\propto n_{\rm s}^{-1}$ (\equnp{Rscrit}). The ratio $\Rs/R_{\rm s,crit}$ is thus expected to increase further as the stream penetrates the halo. We conclude that \textit{the stream will not be disrupted by KHI in the halo}, and the cold gas mass will increase as the stream makes its way towards the central galaxy.

\subsection{Scaling of Properties Within the Halo}
\label{sec:model_scaling}

\smallskip
In this section we describe how different stream properties scale with halocentric radius as the stream penetrates the halo. We assume the stream to be 
on a radial trajectory towards the halo centre and in local pressure equilibrium at every halocentric radius $r$. Based on the results of M20, who showed that variations to the stream temperature induced by the instabilities are very small at all times, we assume the stream to be isothermal throughout the halo. We assume the hot CGM to be isothermal with a power-law density profile of the form $\rho_{\rm h}\propto r^{-\beta}$. Outside of $\sim 0.1\Rv$, this assumption is consistent with cosmological simulations of $\sim 10^{12}\msun$ halos at $z=2$ with $\beta\sim 2$ \citep{vdv12b}, high resolution simulations of isolated halos with $M_{12}\sim 0.1-1$ with $\beta \sim (1.5-2)$ \citep{Fielding17}, and stacked observations of halos with $M_{12}\sim 1-10$ at $z\gsim 0.1$ with $\beta\sim 1.2$ \citep{Singh18}. Analytic models for density profiles of the hot CGM in galaxy clusters predict slopes of $\beta\sim 3$ near the outskirts of the halo, similar to the underlying NFW halo profile \citep{Komatsu01}. Based on these considerations, in what follows we consider values of $\beta=(1-3)$.

\smallskip
\nmr{The stream line-mass is given by}
\be 
\label{eq:m0_model}
\nmr{m(r) = \pi r_{\rm s}^2(r) \rho_{\rm s}(r).}
\ee 
{\no}\nmr{Upon entering the halo, prior to any mass entrainment, we have }
\be 
\label{eq:m0_model_2}
\begin{array}{c}
\nmr{m_0 = \pi\Rs^2 \rho_0 \simeq 10^{10}\msun\kpc^{-1}M_{12}^{2/3}\,(1+z)_3^{-2}}\\
\\
\nmr{\times \Rsv^2\,n_{\rm s,0.01},}
\end{array}
\ee 
{\no}\nmr{where $\Rs=r_{\rm s}(\Rv)$ and we have used \equ{Rvir}.} 

\smallskip
If both the stream and the halo are isothermal and in local pressure equilibrium, the density contrast, $\delta$, remains  constant throughout the halo. In this case, the stream density obeys 
\be 
\label{eq:ns_scaling}
n_{\rm H,s}(r) = \n0\left(\frac{r}{\Rv}\right)^{-\beta}.
\ee
{\no}\nmr{Combining \equs{m0_model} and \equm{ns_scaling} we obtain for the radius of the stream}
\be 
\label{eq:rs_scaling}
\nmr{r_{\rm s}(r) = \Rs \left(\frac{r}{\Rv}\right)^{\beta/2}~\left(\frac{m(r)}{m_0}\right)^{1/2},}
\ee
{\no}\nmr{As we will see in \se{model_sol} below, the stream line mass increases by less than a factor of $\sim 2$ from $\Rv$ to $0.1\Rv$.} Thus, as the stream approaches the center of the halo it gets denser and narrower.


\smallskip
Under our isothermal assumption, the sound speed in both the stream and the background remain constant throughout the halo, 
\be 
\label{eq:cs_cb_scaling}
c_{\rm s}(r) = \cs = {\rm const},\:\:c_{\rm h}(r) = c_{\rm h} = {\rm const}.
\ee
{\no}The stream sound crossing time thus scales as 
\be 
\label{eq:tsc_scaling}
\tsc(r) = \tsc(\Rv) \left(\frac{r}{\Rv}\right)^{\beta/2}~\nmr{\left(\frac{m(r)}{m_0}\right)^{1/2}},
\ee
{\no}with 
\be 
\label{eq:tsc_model}
\tsc(\Rv)=2\frac{\Rs}{\cs} = 2\Rsv\,\delta^{1/2}\,\Mb\,\tv,
\ee 
{\no}where 
\be 
\label{eq:tv_model}
\tv=\frac{\Rv}{V_0} = 0.5\Gyr~\eta^{-1}\,(1+z)_3^{-3/2},
\ee 
{\no}is the halo crossing time of the stream. The cooling time, in both the stream and the mixing region, scales as 
\be 
\label{eq:tcool_scaling}
t_{\rm cool}(r) = t_{\rm cool}(\Rv) \left(\frac{n_{\rm H,s}}{\n0}\right)^{-1} = t_{\rm cool}(\Rv) \left(\frac{r}{\Rv}\right)^{\beta}.
\ee
{\no}Combining \equs{tent}, \equm{tsc_scaling}, and \equm{tcool_scaling}, the entrainment time at radius $r$ is 
\be 
\label{eq:tent_scaling}
t_{\rm ent}(r) = t_0 \left(\frac{r}{\Rv}\right)^{5\beta/8}~\nmr{\left(\frac{m(r)}{m_0}\right)^{3/8}},
\ee
{\no}with 
\be 
\label{eq:t0_model}
t_0\equiv t_{\rm ent}(\Rv) \simeq 0.2 \Rsv\,\delta^{3/2}\,\Mb\,\tau_{\rm cool}^{1/4}\,\tv,
\ee
{\no}where $\tau_{\rm cool}=[t_{\rm cool,1.5\Ts}(\Rv)/\tsc(\Rv)]/0.002$. This normalization is motivated by Table 1 in M20, extrapolated to $\n0=0.01\cmc$ and $\Rs=12\kpc$, which are typical values for $\Mv\gsim 10^{12}\msun$ and $z\gsim 2$ (\fig{params}). 


\subsection{Equations of Motion}
\label{sec:model_motion}

\smallskip
In M20 we derived equations describing the mass entrainment and subsequent stream deceleration induced by the KHI when no external forces are present, which are summarized in \se{theory}. 
We here assume that the same mass entrainment equation can be applied locally at each radius $r$ within the halo. Note that this is \textit{different} than the assumption made in \citet{P18} and \citet{M19}. In those papers, we assumed that KHI introduces an effective drag force with a local deceleration rate identical to the case with no external forces, i.e. the time derivative of \equ{decel_rad}. The net acceleration was then given by the sum of the gravitational induced acceleration and this local deceleration. As will become evident below, these two assumptions lead to different equations of motion, and we find the current assumption to be more physical. Differentiating \equ{mass_rad_approx} yields 
\be 
\label{eq:mdot_model}
{\dot {m}} = \frac{m_0}{t_{\rm ent}(r)}.
\ee 

\smallskip
The stream is subject to gravitational \nmr{acceleration} by the external halo. The external force acting on the stream per unit length, including the entrained material, is 
\be 
\label{eq:Fext}
\mathcal{F_{\rm ext}} = -m(r) \frac{{\rm d}\Phi}{{\rm d}r},
\ee
{\no}where $\Phi(r)$ is the gravitational potential of the dark matter halo at $r$. For an NFW halo with concentration parameter $c$, 
\be 
\label{eq:Phi_ext}
\Phi = -\Vv^2\frac{{\rm ln}\left(1+cx\right)}{f_c~x},
\ee
{\no}where
\be 
\label{eq:x_def}
x\equiv \frac{r}{\Rv},
\ee
{\no}and
\be 
\label{eq:f_def}
f_c \equiv {\rm ln}\left(1+c\right) - \frac{c}{1+c}.
\ee

\smallskip
We define $\mathcal{P}=m(r)V(r)$ as the total momentum per unit length of stream including the entrained material. Using \equ{mdot_model} and the fact that ${\dot {V}}=V(r){\rm d}V/{\rm d}r$, we have 
\be 
\label{eq:Pdot}
{\dot {\mathcal{P}}} = m(r)\left(\frac{1}{t_{\rm ent}(r)}\frac{m_0}{m(r)}V(r) + \frac{1}{2}\frac{{\rm d}V^2}{{\rm d}r}\right),
\ee 
{\no} The equation of motion for the stream is then 
\be 
\label{eq:eom}
{\dot {\mathcal{P}}} = \mathcal{F_{\rm ext}}.
\ee

\smallskip 
Defining dimensionless variables  
\be 
\label{eq:y_def}
y\equiv \frac{V^2}{\Vv^2},\:\:\phi\equiv \frac{\Phi}{\Vv^2},\:\:\mu\equiv \frac{m}{m_0},\:\:\tau\equiv\frac{\Rv}{V_0 t_0}=\frac{\tv}{t_0},
\ee 
{\no}\equs{tent_scaling} and \equm{Fext}-\equm{y_def} can be combined to yield an equation for the velocity profile within the halo\footnote{Had we followed the different, less physically motivated, assumption of \citet{P18} and \citet{M19} described above, \equ{y_prime} would be replaced by ${\rm d}y/{\rm d}x=2\tau y\,x^{-5\beta/8} - 2{\rm d}\phi/{\rm d}x$.},
\be 
\label{eq:y_prime}
\frac{{\rm d}y}{{\rm d}x} = \frac{2 \, \eta \, \tau \, y^{1/2}}{\nmr{\mu^{11/8}} \, x^{5\beta/8}} - 2\frac{{\rm d}\phi}{{\rm d}x},
\ee
{\no}with the boundary condition that $y(x=1)=\eta^2$. We have used the sign convention that the inward velocity is negative, such that $V/\Vv=-y^{1/2}$. If there were no KHI induced mass entrainment, then $t_0\rightarrow \infty$, $\tau=0$, and \equ{y_prime} represents gravitational free-fall. 

\smallskip
Using the relation ${\dot {m}}=V(r){\rm d}m/{\rm d}r$, \equs{mdot_model} and \equm{y_def} can be combined to yield an equation for the stream mass profile within the halo, including the entrained background mass, 
\be 
\label{eq:mu_prime}
\frac{{\rm d}\mu}{{\rm d}x} = -\frac{\eta \, \tau}{y^{1/2}\,x^{5\beta/8}\,\nmr{\mu^{3/8}}},
\ee 
{\no}with the boundary condition $\mu(x=1)=1$.

\smallskip
Using \equ{t0_model}, the characteristic time scale in these two equations is given by
\be 
\label{eq:tau_model_1}
\tau \simeq 5 \Rsv^{-1}\,\delta^{-3/2}\,\Mb^{-1}\,\tau_{\rm cool}^{-1/4}.
\ee
{\no}For a $10^{12}\msun$ halo at $z=2$, with our fiducial values of $\delta\sim 100$ (\equnp{delta_model}) and $\Rsv\sim 0.16$ (\equnp{Rs_Rv_model}), $\tau \sim 0.03$ for $\Mb\sim \tau_{\rm cool}\sim 1$. We thus have $t_{\rm 0}\gg t_{\rm v}$ (\equnp{y_def}), so the virial crossing time is much less than the mass entrainment time at $\Rv$. In this case, the stream mass is not expected to increase significantly before reaching the central galaxy. This is consistent with the notion of constant mass flux along the stream discussed above, and is also consistent with model solutions shown in \se{model_sol} below.

\subsection{Energy Dissipation and Radiation}
\label{sec:model_emis}

\smallskip
In the case with no external forces studied in M20, the energy dissipation is given by the dissipation of kinetic and thermal energy (\se{theory}). In this case, however, we must consider dissipation of \textit{mechanical} (kinetic plus potential) and thermal energy. We begin with the mechanical energy of our system. By ``our system'', we here mean the initial stream plus all the mass that will become entrained in the stream before it reaches the central galaxy. Accounting for this extra mass from the outset is important, since it has initial potential energy before becoming entrained in the stream. The mechanical energy per unit length of the system at radius $r$ is thus 
\be 
\label{eq:Emech}
\begin{array}{c}
E_{\rm mech}(r)=0.5 m(r)V(r)^2 + m(r)\Phi(r) \\
\\
+ {\displaystyle \int_{R_{\rm int}}^{r}} \left|\dfrac{{\rm d}m}{{\rm d}\tilde{r}}\right|\Phi(\tilde{r}) {\rm d}\tilde{r},
\end{array}
\ee
{\no}where $R_{\rm int}$ is the innermost radius we are considering. The final term represents the potential energy of material not yet entrained by the stream at $r$, but which will become entrained before the stream reaches $R_{\rm int}$. Note that ${\rm d}m/{\rm d}r<0$.  

\smallskip
Using ${\dot {E}}=V{\rm d}E/{\rm d}r$ together with \equs{x_def}, \equm{y_def}-\equm{mu_prime} and \equm{Emech}, we obtain the dissipation rate of mechanical energy per unit length 
\be 
\label{eq:Edot_mech}
{\dot {E}}_{\rm mech}(r) = -\frac{m_0\Vv^2}{2 t_0}~\frac{y}{ x^{5\beta/8}\,\nmr{\mu^{3/8}} }.
\ee
{\no}Note that there is no explicit dependence on the potential, except through the solution to \equ{y_prime} for $y(x)$. With no KHI induced mass entrainment, $t_0\rightarrow\infty$ and ${\dot {E}}_{\rm mech}=0$ as expected.

\smallskip
The dissipation rate of thermal energy is governed by the entrainment rate of background mass onto the stream, which we have assumed to be locally the same as in M20. Therefore, we can use \equ{erad} to obtain 
\be 
\label{eq:Tdot_model}
\dot{E}_{\rm th}(r) = -\frac{9m_0 V_0^2}{10\Mb^2t_{\rm ent}(r)} \simeq -\frac{m_0\Vv^2}{2t_0}\frac{1.6\Theta_{\rm h}}{x^{5\beta/8}\,\nmr{\mu^{3/8}}},
\ee
{\no}with $V_0=\eta\Vv$ and $\eta\,\Mb^{-1}\sim 0.9\Theta_{\rm h}^{1/2}$ from \equ{Mb_isothermal}.

\smallskip
Generalizing \equ{Ldiss}, the total radiation emitted between radius $r$ and $\Rv$ is
\be 
\label{eq:Ldiss_model_2}
L_{\rm diss}(>r) = \int_{r}^{\Rv} \left|{\dot {E}}_{\rm mech}(\tilde{r})\right| + \frac{5}{3}\left|{\dot {E}}_{\rm th}(\tilde{r})\right| {\rm d\tilde{r}}.
\ee

\begin{figure*}
\begin{center}
\includegraphics[trim={0.02cm 0.09cm 0.01cm 0.01cm}, clip, width =0.99 \textwidth]{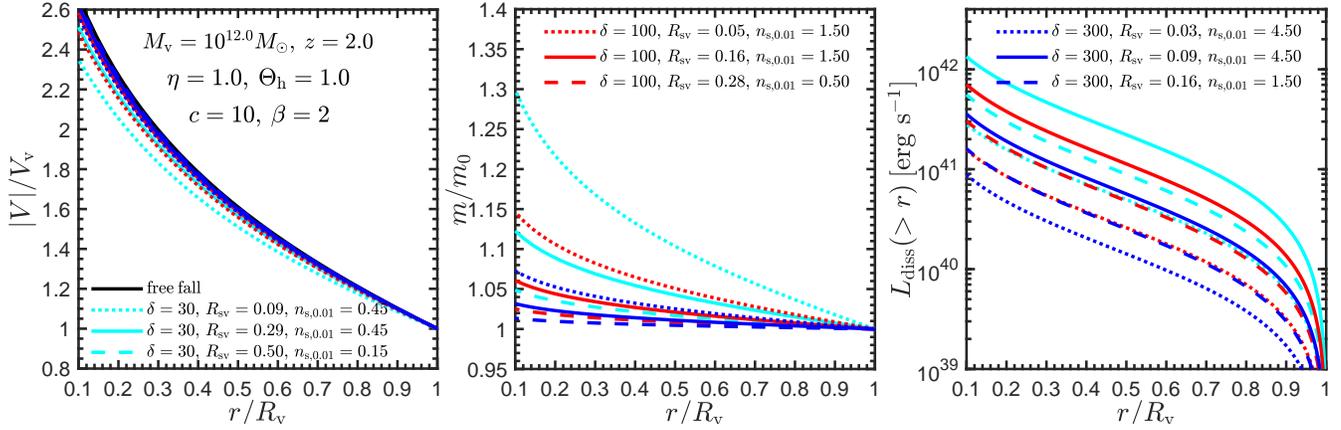}
\end{center}
\caption{Solutions of our toy model for stream evolution, in an NFW halo with virial mass $\Mv=10^{12}\msun$ and concentration $c=10$ at redshift $z=2$, with a CGM density slope of $\beta=2$. Cyan, red, and blue lines show solutions for $\delta=30$, 100, and 300 respectively. Different line styles represent different values of the stream density, $n_{\rm s,0.01}=n_0/(0.01\cmc)$, and the stream radius normalized by the virial radius, $R_{\rm sv}=\Rs/\Rv$, as indicated in the legend. For each $\delta$, these correspond to values of the model parameters $\widetilde{f}_{h}=(1-3)$ and $\widetilde{s}=(0.3-3)$, and are near the upper and lower bounds expected for these parameters for this halo mass and redshift (\fig{params}). Dotted lines represent dense and narrow streams, solid lines represent dense and wide streams, and dashed lines represent dilute and wide streams. All models assume $\eta=\Theta_{\rm h}=1$. In each panel, the x axis is the halocentric radius normalized to the halo virial radius.  \textit{On the left}, we show the stream velocity normalized to the virial velocity, with the black line representing the free-fall velocity profile if there were no KHI or mass entrainment, $V_{\rm ff}$. All cases undergo net acceleration, though the deceleration with respect to free-fall increases with lower $\delta$, narrower and more dilute streams. However, the stream velocity at $0.1\Rv$ is at least $\sim 0.8V_{\rm ff}$ for the range of parameters considered here. \textit{In the centre}, we show the total line-mass, including the background mass entrained in the stream. Lower values of $\delta$, as well as narrower and more dilute streams, entrain more background mass. At $0.1\Rv$, the stream can increase its line-mass by up to $\sim 35\%$. \textit{On the right}, we show the luminosity induced by (mechanic plus thermal) energy dissipation between each radius $r$ and $\Rv$. KHI induced dissipation can emit $\sim (10^{41}-10^{42}){\rm erg\,s^{-1}}$ within the halo, with $\sim 90\%$ of the dissipation occurring in the inner $\lsim 0.6\Rv$. }
\label{fig:toy_model} 
\end{figure*}

\smallskip 
Using \equs{Vvir}, \equm{tv_model}, \equm{t0_model}, \equm{m0_model_2}, and \equm{Ldiss_model_2}, the normalization of the total radiation emitted between radius $r$ and $\Rv$ is  
\be 
\label{eq:Ldiss_norm}
\begin{array}{c}
\dfrac{m_0 \Rv \Vv^2}{2 t_0}\simeq 2\times 10^{40}{\rm erg\,s^{-1}}~M_{12}^{5/3}(1+z)_3^{-1/2}\\
\\
\times R_{\rm sv,0.16}\,n_{\rm s,0.01}\,\delta_{100}^{-3/2}\,\Theta_{\rm h}^{1/2}\,\tau_{\rm cool}^{-1/4},
\end{array}
\ee
{\no}with $R_{\rm sv,0.16}=\Rsv/0.16$. \nmr{Together with \equs{delta_model}, \equm{n0}, and \equm{Rs_Rv_model}, we obtain that the normalization scales as $\Mv\,(1+z)^2$. In practice, the emitted radiation, $L_{\rm diss}$, deviates from this scaling due to the dependence of $\mu$ and $y$ in \equs{Edot_mech} and \equm{Tdot_model} on halo mass and redshift.}

\smallskip
We stress that \equ{Ldiss_model_2} represents the bolometric luminosity emitted from a single stream as a result of the interaction between the stream and the ambient hot CGM. As a typical halo is fed by $\sim 3$ streams \citep[e.g.][]{Dekel09}, the bolometric luminosity emitted from the halo as a result of stream-CGM interaction will be $\sim 3$ times larger. Roughly half of this luminosity is expected to be emitted from gas with $T\sim (1-5)\times 10^4\K$ (M20), and $\gsim$ half of this will be emitted in Ly$\alpha$ through collisional excitation \citep[e.g.][]{Katz96,Fardal01,Goerdt10}. We thus expect the total Ly$\alpha$ luminosity emitted from the halo as a result of stream-CGM interaction to be comparable to the estimate given by \equ{Ldiss_model_2}. However, excluded from this estimate is Ly$\alpha$ emission resulting from fluorescence of UV background photons. This can increase the total Ly$\alpha$ luminosity by factors of up to a few, depending on the details of the stream density structure and self-shielding (M20). We ignore this contribution for now, postponing it's treatment to future work which will incorporate realistic self-shielding in simulations. We therefore treat \equ{Ldiss_model_2} as an approximate lower limit to the total Ly$\alpha$ emission from halos fed by cold streams.

\subsection{Example Solutions}
\label{sec:model_sol}

\subsubsection{$\beta=2$}

\smallskip
We here present solutions of our toy model for a fiducial CGM density slope of $\beta=2$, addressing $\beta$ dependence below. 
\Fig{toy_model} presents the radial profiles of velocity (left), line-mass (centre), and luminosity (right) of a single stream, resulting from the stream-CGM interaction according to our model. We show results for a $10^{12}\msun$ halo at $z=2$ with concentration $c=10$. Cyan, red, and blue curves show solutions for $\delta=30$, 100, and 300 respectively. Different line styles show different values of $n_{\rm s,0.01}$ and $\Rsv$, obtained from \equs{n0} and \equm{Rs_Rv_model} using $[\widetilde{f}_{\rm h},\widetilde{s}]=[3,0.3]$ (dotted lines), $[3,3]$ (solid lines), and $[1,3]$ (dashed lines) with the given values of $\delta$ and $z$. The corresponding values of $\Rsv$ and $n_{\rm s,0.01}$ are listed in the legend, and are near the upper and lower bounds of the parameter ranges shown in \fig{params} for $\Mv\sim 10^{12}\msun$ and $z\sim 2$. These three models represent a narrow and dense stream, a wide and dense stream, and a wide and dilute stream respectively. All models assume $\Theta_{\rm h}=\eta=1$, and the initial stream temperature is determined by $\delta$. We evaluate the cooling time $\tau_{\rm cool}$ in \equ{t0_model} assuming a \nmr{metallicity} of $Z_{\rm s}=0.03Z_{\odot}$, though our results are not strongly affected by this choice so long as $Z_{\rm s}\lsim 0.1Z_{\odot}$. 

\smallskip
\nmr{For these parameters, the stream line mass is $m_0 \sim 10^{10}\msun\kpc^{-1}\Rsv^2\,n_{\rm s,0.01}$ (\equnp{m0_model_2}). The virial radius and velocity are $\Vv\sim 200\kms$ and $\Rv\sim 100\kpc$. For typical values of $\Rsv\sim 0.16$ and $n_{\rm s,0.01}\sim 1.5$ (\fig{toy_model}), the total stream mass is thus $M_{\rm stream}\sim m_0\Rv\sim 3\times 10^{10}\msun$. Assuming all the cold gas in the CGM is contained in three such streams, our model predicts a cold CGM gas mass of $M_{\rm cold}\sim 10^{11}\msun$ in halos of $\Mv\sim 10^{12}\msun$ at redshift $z\sim 2$. This is within a factor of $<2$ from recent observational estimates of the cold gas content in the CGM of halos with similar masses and redshifts, which host giant Lyman-$\alpha$ nebulae \citep{Pezzulli19}.}

\smallskip
The velocity profiles show that in all cases the stream accelerates towards the halo centre, with the stream velocity only slightly lower than the free-fall velocity had there been no mass entrainment or KHI, shown by the solid black line. The deceleration with respect to free-fall is greater for smaller values of $\delta$, and for narrower and denser streams. However, for the range of parameters explored here, the stream velocity at $r=0.1\Rv$ is at least $\sim 0.8$ times the free-fall velocity. 
\nmr{Overall, the velocity increases by a factor of $\sim (2-2.5)$ from $\Rv$ to $0.1\Rv$.}
\nmr{Note that our current model predicts significantly less deceleration than \citet{M19}.} This is not due to the current consideration of radiative cooling effects, but rather to the different assumption made in deriving our equation of motion (\equnp{y_prime}), as discussed in \se{model_motion}.

\smallskip
As expected from the velocity profiles, the entrained mass is larger for narrower, denser streams, with smaller density contrasts. 
\nmr{For the range of parameters shown, the stream mass increases by $\lsim 30\%$ from $\Rv$ to $0.1\Rv$ for this halo mass and redshift.} 

\begin{figure*}
\begin{center}
\includegraphics[trim={0.02cm 0.02cm 0.01cm 0.01cm}, clip, width =0.99 \textwidth]{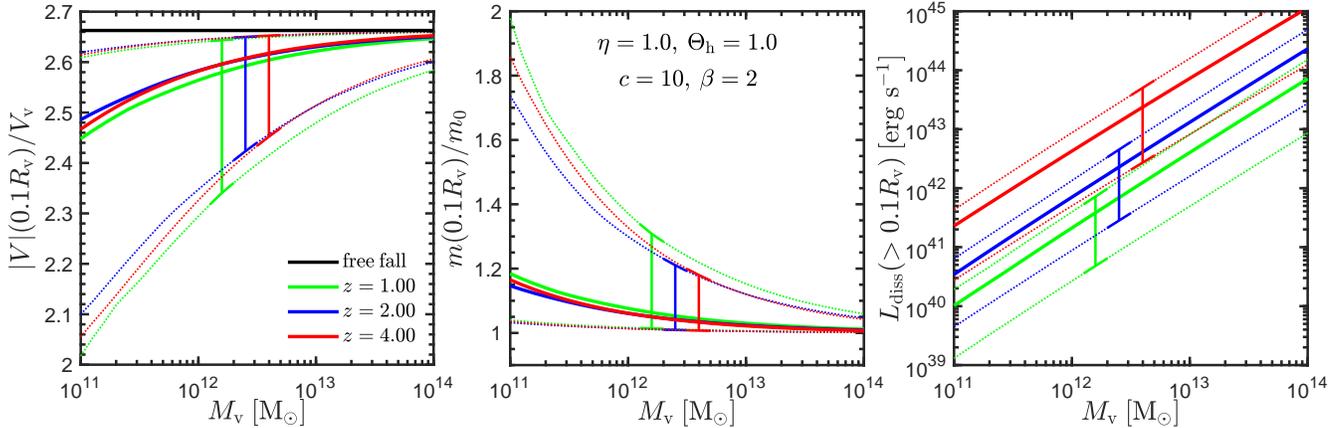}
\end{center}
\caption{Stream properties at $0.1\Rv$ according to our model, as a function of halo mass and redshift. We show the stream velocity (left), the line-mass (centre), and the total luminosity induced by the stream-CGM interaction (right). Different colours and line-styles are as in \figs{params} and \figss{Rscrit}. The range of model parameters at $\Mv=10^{12}\msun$ and $z=2$ is identical to those used in \fig{toy_model}. For different values of $\Mv$ and $z$, $\delta$, $n_{\rm s,0.01}$, and $\Rsv$ were then scaled according to \equs{delta_model}, \equm{n0}, and \equm{Rs_Rv_model} respectively. Streams in more massive halos entrain less mass from the CGM compared to their initial mass, and thus decelerate less compared to the free-fall velocity. For $\Mv\sim 10^{11}\msun$, the stream mass can potentially double, while for $\Mv\sim 10^{14}\msun$ it increases by $\lsim 5\%$ at most, roughly independent of redshift. Streams always net-accelerate towards the halo centre, with the velocity at $0.1\Rv$ at least $\sim 70\%~(90\%$) of the free fall velocity for $10^{11}~(10^{14})\msun$ halos. The emitted luminosity is a strong function of both halo mass and redshift (\equnp{Ldiss_norm}). Luminosities greater than $10^{43}~{\rm erg\,s^{-1}}$ are possible in halos with $\Mv \gsim 10^{13}$, $5\times 10^{12}$, and $10^{12}\msun$ at $z=1$, $2$, and $4$ respectively.}
\label{fig:Ltot_panels} 
\end{figure*}

\smallskip
The luminosity profiles show that the emission is highest for low values of $\delta$, in line with the expectation from \equ{Ldiss_norm}, where the luminosity increases with decreasing $\delta$ and increasing $\Rsv$ and $n_{\rm s,0.01}$. Intuitively, the increased emission with stream radius is due to the larger stream surface area and resulting increased interaction with background gas, while the increased emission with stream density is due to the decreased cooling time at higher densities. The increased emission with lower density contrast stems from the entrainment time being shorter for lower values of density contrast (\equnp{t0_model}), and is consistent with more mass entrainment and deceleration. For the range of parameters shown here, the total luminosity produced by a single stream in a $10^{12}\msun$ halo at $z=2$ with CGM density slope $\beta=2$ is in the range $\sim (10^{41}- 10^{42})~{\rm erg\,s^{-1}}$, with $50\%\,(90\%)$ of the luminosity emitted within $\sim 0.3\Rv\,(0.6\Rv)$.


\smallskip
The results for halo concentrations of $c=5$ and $20$ are extremely similar. The only noticeable difference is that the free-fall velocity increases with halo concentration. However, the deceleration with respect to free-fall, the mass entrainment, and the luminosity do not depend strongly on $c$.

\smallskip
\Fig{Ltot_panels} shows stream properties at $0.1\Rv$ predicted by our model, as a function of halo mass and redshift. We show the stream velocity normalized by the virial velocity (left), the stream line-mass normalized by the initial line-mass (centre), and the total emitted luminosity (right). At each halo mass and redshift, we explore a range of values for $\delta$, $n_{\rm s,0.01}$, and $\Rsv$. These were normalized for a $10^{12}\msun$ at $z=2$ to the same values used in \fig{toy_model}, and then scaled to different halo masses and redshifts following \equs{delta_model}, \equm{n0}, and \equm{Rs_Rv_model}. We thus solve nine models for each value of $\Mv$ and $z$, all of which assume $\Theta_{\rm h}=\eta=c=1$ and $\beta=2$ as in \fig{toy_model}, and plot the resulting range of solutions, as in \figs{params} and \figss{Rscrit}. The thick lines represent a model which has $(\delta,\Rsv,n_{\rm s,0.01})=(100,0.16,1.50)$ for $M_{12}=(1+z)_3=1$ (solid red lines in \fig{toy_model}). 

\smallskip
As the halo mass increases, the stream entrains less mass and its velocity approaches the free-fall velocity. For $\Mv=10^{11}\msun$, the stream line-mass can grow by up to a factor of 2 from $\Rv$ to $0.1\Rv$, while the velocity at $0.1\Rv$ can be as low as 0.7 times the free-fall value. For a $10^{14}\msun$ halo, the stream mass increases by $\lsim 5\%$, and the velocity is $\gsim 95\%$ of the free-fall value. This is primarily due to the increase of $\delta$ with $\Mv$, resulting in a longer entrainment timescale (\equnp{t0_model}), and is roughly independent of redshift. 

\begin{figure}
\begin{center}
\includegraphics[trim={0.01cm 0.03cm 0.01cm 0.01cm}, clip, width =0.48 \textwidth]{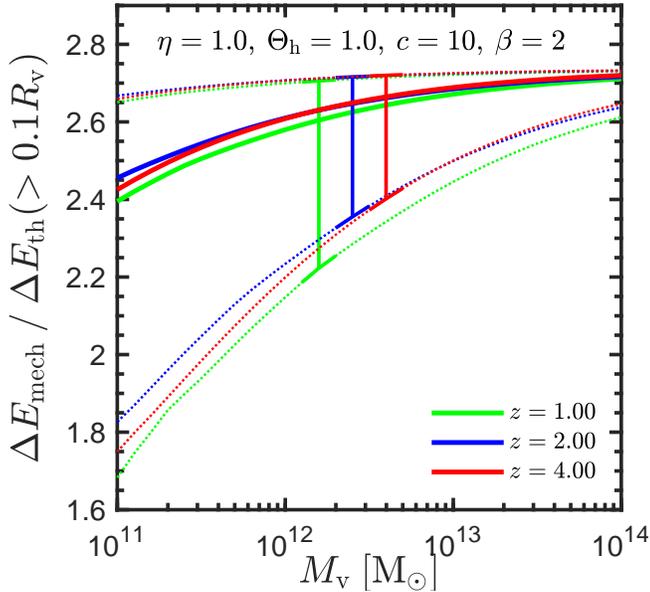}
\end{center}
\caption{Ratio of dissipated mechanical to thermal energy from $\Rv$ to $0.1\Rv$, as a function of halo mass and redshift. Line styles and colours are the same as in \fig{Ltot_panels}, as are the model parameters used. Roughly twice as much mechanical energy as thermal energy is lost at all redshifts, though the ratio decreases slightly towards lower halo masses. 
}
\label{fig:Lrat} 
\end{figure}

\begin{figure*}
\begin{center}
\includegraphics[trim={0.02cm 0.02cm 0.01cm 0.01cm}, clip, width =0.99 \textwidth]{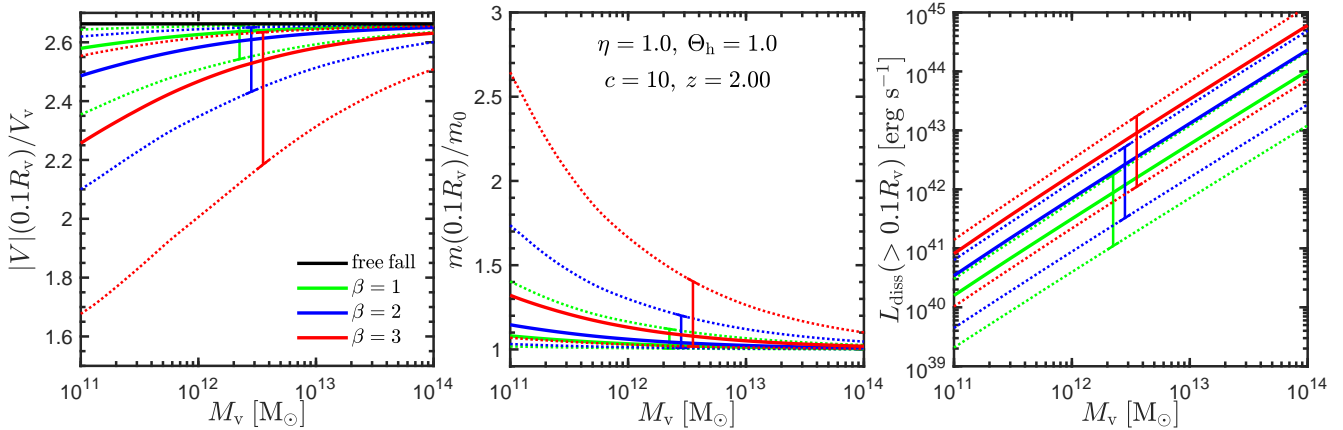}
\end{center}
\caption{Similar to \fig{Ltot_panels}, we show as a function of halo mass the stream velocity (left), the line-mass (centre), and the total luminosity induced by the stream-CGM interaction (right) at $0.1\Rv$. The redshift is kept fixed at $z=2$, while green, blue, and red lines represent $\beta=1$, $2$, and $3$ respectively. All other model parameters are as in \fig{Ltot_panels}, as are the meanings of the different linestyles. As $\beta$ increases, the inner CGM becomes denser, leading to increased mass entrainment, deceleration, and emission from streams. The trends with halo mass and redshift are qualitatively similar for all $\beta=(1-3)$.}
\label{fig:Ltot_panels_beta} 
\end{figure*}

\smallskip
The total luminosity emitted at $r>0.1\Rv$ increases with both halo mass and redshift, in accordance with \equ{Ldiss_norm}. Luminosities of $L_{\rm diss}\gsim 10^{43}~{\rm erg\,s^{-1}}$ can be reached in halos with $\Mv/\msun \gsim 10^{13}$, $4\times 10^{12}$, and $10^{12}$ at $z=1$, $2$, and $4$ respectively, while $L_{\rm diss}>10^{42}~(10^{44})~{\rm erg\,s^{-1}}$ can be achieved in halos $\sim 5$ times less (more) massive. As discussed at the end of \se{model_emis}, we expect this to be comparable to the total Ly$\alpha$ luminosity produced by the stream-CGM interaction, and a rough lower limit to the total Ly$\alpha$ luminosity emitted by all sources in the halo. As described in \se{intro}, observed LABs range from a few times $10^{42}$ to $\sim 10^{44}{\rm erg\,s^{-1}}$, and extend on the sky to $\gsim 50\kpc$ or more, $\gsim 0.5\Rv$ for a $\gsim 10^{12}\msun$ halo at $z\gsim 2$. Our model thus suggests that the interaction between cold streams and a hot CGM is certainly a viable source for low-to-intermediate luminosity LABs, and potentially for high-luminosity LABs in $\gsim 10^{13}\msun$ halos at $z\gsim 3$. 

\smallskip
\Fig{Lrat} shows the ratio of the mechanical energy dissipated from $\Rv$ to $0.1\Rv$ to the thermal energy dissipated in the same interval, for the same parameters used in \fig{Ltot_panels}. The ratio is nearly independent of redshift and depends only weakly on halo mass. For $10^{11}\msun$ halos, the system dissipates $\sim (1.6-2.7)$ times more mechanical than thermal energy. For $10^{14}\msun$ halos, the lost mechanical energy is $\sim (2.5-2.75)$ times greater than the lost thermal energy. In M20, we found that at $t\ll t_{\rm ent}$, $\Delta E_{\rm k}/\Delta E_{\rm th}\sim 0.5\Mb^2$. Substituting $\Delta E_{\rm mech}$ for $\Delta E_{\rm k}$, our current results are consistent with this, since most of the energy is dissipated in the inner halo where the stream velocity is a factor of $\sim 2$ larger than the initial velocity, and thus $\Mb\sim 2$ (\fig{toy_model}). 

\subsubsection{$\beta$ dependence}

\smallskip
\Fig{Ltot_panels_beta} shows the stream properties at $0.1\Rv$ predicted by our model as a function of halo mass for $\beta=1$ (green), $2$ (blue) and $3$ (red). We show only the results for redshift $z=2$ (blue lines in \fig{Ltot_panels}), and use the same values of $c$, $\eta$, $\delta$, $n_{\rm s,0.01}$, and $\Rsv$ used in \fig{Ltot_panels}. Since $n_{\rm s,0.01}$ normalizes the density at $\Rv$, larger values of $\beta$ imply larger gas densities in the inner halo. Thus, as $\beta$ increases, the stream entrains more mass and decelearates more with respect to free-fall. For $\Mv=10^{11}\msun$, the stream line-mass can grow from $\Rv$ to $0.1\Rv$ by factors of up to 1.5 and 4 for $\beta=1$ and 3, respectively, while the corresponding velocities at $0.1\Rv$ can be as low as 0.87 and 0.55 times the free-fall value. For a $10^{14}\msun$ halo, the stream mass increases by $\lsim 5\%$, and the velocity is $\gsim 95\%$ of the free-fall value for $\beta=(1-3)$. These results are rather insensitive to redshift, as seen in \fig{Ltot_panels} for $\beta=2$.

\smallskip
The total luminosity emitted at $r>0.1\Rv$ increases with $\beta$, as a result of the increased mass entrainment and deceleration. While we only show results for $z=2$, the scaling with redshift for all $\beta$ is similar to that seen for $\beta=2$ in \fig{Ltot_panels}, and follows from \equ{Ldiss_norm}. For $\beta=3$, luminosities of $L_{\rm diss}\gsim 10^{43}~{\rm erg\,s^{-1}}$ can be reached in halos with $\Mv/\msun \gsim 5\times 10^{12}$, $2\times 10^{12}$, and $5\times 10^{11}$ at $z=1$, $2$, and $4$ respectively, while $L_{\rm diss}>10^{42}~(10^{44})~{\rm erg\,s^{-1}}$ can be achieved in halos $\sim 5$ times less (more) massive. For $\beta=1$, these threshold masses increase by a factor of $\sim 4$. The radial profiles of the emission become steeper with increasing $\beta$, reflecting the steepening of the CGM density profile. As $\beta$ increases from $1$ to $3$, the radius containing $50\%\,(90\%)$ of the total emission outside $0.1\Rv$ decreases from $\sim 0.4\Rv\,(0.8\Rv)$ to $\sim 0.2\Rv\,(0.45\Rv)$. 

\section{Limitations of the Current Model}
\label{sec:caveat} 

\smallskip
It is encouraging that our model, based on the dissipation mechanisms described in detail in M20, can produce spatially extended emission with luminosities comparable to observed LABs in halos of relevant mass and redshift. Our current model presents a significant improvement over those presented in \citet{P18} and \citet{M19}. Furthermore, unlike the analytic models presented in \citet{Dijkstra09} and \citet{Goerdt10}, our model invokes a well-defined dissipation mechanism to produce the radiation. However, it is still very simplified and limited in a number of ways, which we discuss here. 

\smallskip
Firstly, we have only addressed the bolometric luminosity produced by the instability, not specifically the Ly$\alpha$ emissivity. As highlighted at the end of \se{theory} and \se{model_emis}, the simulations presented in M20 found that roughly half of the radiation is emitted from gas with temperatures $T\sim (1.5-5)\times 10^4\K$, and is thus likely to be dominated by Ly$\alpha$. When accounting for the fact that halos typically have $\sim 3$ streams, we estimated that the single-stream luminosities computed in our model are comparable to the total Ly$\alpha$ luminosity produced by stream-CGM interactions in the halo. However, the simulations of M20 did not include self-shielding from the UV background. While this is unlikely to alter the total dissipated energy or bolometric luminosity, it may alter the temperature distribution of the emitting gas, and thus the Ly$\alpha$ contribution \citep{FG10}. However, since most of the emission comes from the turbulent mixing zone surrounding the stream rather than from the stream interior (M20, see also \citealp{Gronke20}), it is unclear whether this gas will be strongly self-shielded. We therefore speculate that a significant fraction of the emission will indeed be in Ly$\alpha$. We will address this issue in future work, where we will include self-shielding in our simulations.

\smallskip
Related to the previous point, and as also highlighted at the end of \se{model_emis}, we have only computed a lower limit to the luminosity, produced purely by the stream-CGM interaction. As shown by M20 \citep[see also][]{Goerdt10,FG10}, fluorescent radiation caused by the UV background can contribute significantly to the total luminosity. The magnitude of this contribution is unclear and depends on the level of self-shielding as discussed above, but this does imply that the Ly$\alpha$ emission from cold streams may be even larger than predicted here. We will consider this along with self-shielding in future work.

\smallskip
The radial profiles of the emission predicted by our model are in reasonable agreement with that found for the cooling radiation from the CGM of a $\sim 10^{12}\msun$ halo at $z\sim 3$, found in a cosmological simulation with full radiative transfer \citep{Trebitsch16}. However, as described in that paper, the radial profiles of the observed emission may be much flatter due to scattering of the radiation from the inner halo to larger radii, which also leads to polarized emission as observed \citep{Trebitsch16}. Therefore, we cannot address the observed emission profiles until accounting for radiative transfer effects, which we defer to future work.

\smallskip
In~\se{model_scaling},~we~made~a~number of simplifying assumptions~when evaluating the scaling of stream properties~with~halocentric~radius. First, the assumption that the hot CGM is isothermal with a power-law density profile is clearly an oversimplification. This was motivated by cosmological and isolated simulations and low-$z$ observations \citep{vdv12b,Fielding17,Singh18}, and makes the model analytically tractable. As the properties of the high-$z$ CGM are not observationally constrained, we are confident that the range of profiles we considered likely brackets realistic values for the resulting luminosities. \nmr{However, when better constraints for the high-$z$ CGM are available, the model should be refined to reflect these.} 

\smallskip
Furthermore, we assumed the streams to be on purely radial orbits towards the halo centres. Cosmological simulations suggest that streams have typical impact parameters of $\lsim 0.3\Rv$, which plays a large role in the growth of angular momentum in disc galaxies \citep[e.g.][]{Danovich15}. It has also been suggested that this impact parameter can make the streams unstable to Rayleigh-Taylor instabilities in the inner halo \citep{Keres09}. This is likely to affect our results near the inner halo, which should be addressed in future work incorporating non-radial orbits. However, it seems likely that this will increase the cooling emission, particularly near pericentre where streams are likely to interact and lead to strong shocks \citep[e.g.][]{M18}. Another complication in the inner halo is the interaction of cold streams with outflowing gas from the galaxies. This changes the density and pressure of the background gas in the inner halo, affecting the KHI, and is also likely to lead to strong shocks in the streams, which may disrupt the streams in the inner halo altogether. These shocks are likely to increase the cooling radiation in the inner halo, though they are unlikely to affect our results at $r\gsim (0.2-0.3)\Rv$. Modeling the detailed interaction between inflows and outflows is beyond the scope of this paper, and is left for future work. 

\smallskip
Finally, while our analytic toy model has allowed us to gain insight into the evolution of cold streams in dark matter halos, there may be additional physical effects for which we did not account, such as the tidal field of the halo, which may alter the evolution of KHI. This may invalidate the main assumption, which was that the KHI induced mass entrainment derived by M20 could be applied locally at each halocentric radius. Furthermore, M20 did not account for additional physics, such as self-gravity, magnetic fields, thermal conduction, or cosmic rays\footnote{See section 6 in M20 for a discussion of the potential influence of these effects on their results}, or for initially non-linear perturbations in the streams caused by, e.g., subhalos or mergers along the streams. All of these will be studied in detail in future work both analytically and using idealized simulations, as outlined in M20. Nevertheless, we hope that the model presented here will serve as a useful benchmark against which future simulations that include an explicit halo potential can be compared.

\section{Summary and Conclusions}
\label{sec:conc} 

\smallskip
Massive halos of $\Mv\gsim 10^{12}\msun$ at redshift $z\gsim 2$ are understood to be fed by cold streams with temperatures $\Ts\gsim 10^4\K$. These streams flow along cosmic web filaments and penetrate the hot CGM of these halos, with $T_{\rm h}\gsim 10^6\K$. In our previous paper \citep{M20}, we showed how the interaction between these cold streams and the hot CGM leads to energy dissipation and radiation through the combination of Kelvin-Helmholtz instability (KHI) and radiative cooling. Here, we expanded upon these results by modeling the effect of the dark-matter halo potential with a toy model. In our model, the dark matter halo has three main effects on stream evolution:

\begin{enumerate}
    \item It accelerates the stream towards the halo centre, which counteracts the KHI induced deceleration, and shortens the amount of time a given stream element spends flowing towards the central galaxy.
    
    \item It induces a density profile in both the stream and the CGM gas, such that the gas density and emissivity both grow towards the halo centre.
    
    \item It focuses the stream towards the halo centre, reducing the surface area of the stream, and hence the total emitted radiation, closer to the halo centre.
\end{enumerate}

\smallskip
Combining~this model with cosmologically motivated boundary conditions for the values of stream parameters at $\Rv$ as a function of halo mass and redshift, we estimated the luminosity emitted by a typical cold stream as a result of KHI, as it penetrates the halo towards the central galaxy. We found both the magnitude and spatial extent of this to be comparable to those of observed Ly$\alpha$ blobs (LABs), namely $L_{\rm diss}>10^{42}{\rm erg\,s^{-1}}$ emitted within $\lsim 0.6\Rv$ of halos with $\Mv>10^{12}\msun$ at $z\gsim 2$. This supports previous claims that LABs may constitute direct observational evidence of cold streams \citep{Dijkstra09,Goerdt10}, and expands upon those works by identifying, for the first time, a self-consistent dissipation mechanism powering the emission. \nmr{We predict that the LAB luminosity increases with halo mass slightly super-linearly (\fig{Ltot_panels}). This is similar to the scaling predicted by the model of \citet{Dijkstra09}, which was found to be consistent with observed LAB luminosity functions and clustering.} While our model makes a number of simplifying assumptions, it is encouraging that the dissipation induced by KHI seems to produce emission in the right ball-park for explaining LABs, both in terms of luminosity and spatial extent. Future work should continue to explore these issues using more refined analytic models and idealized simulations that account for additional physics.

\section*{Acknowledgments} 
\nmr{We thank the referee, Andreas Burkert, for his thoughtful comments and constructive report which helped improve the quality and clarity of this manuscript.} 
We thank Nicolas Cornuault, Drummond Fielding, Max Gronke, Joseph F. Hennawi, Suoqing Ji, Neal Katz, S. Peng Oh, X. Prochaska, Santi Roca-Fabrega, and Chuck Steidel for helpful discussions. NM and FCvdB acknowledge support from the Klauss Tschira Foundation through the HITS Yale Program in Astropysics (HYPA). FCvdB received additional support from the National Aeronautics and Space Administration through grant No. 17-ATP17-0028Grant Nos. 17-ATP17-0028 and 19-ATP19-0059 issued as part of the Astrophysics Theory Program. 
DN acknowledges support by National Science Foundation grant AST-1412768 and the hospitality at the Aspen Center for Physics, which is supported by National Science Foundation grant PHY-1607611. AD is
partly supported by the grants NSF AST-1405962,
BSF 2014-273, GIF I-1341-303.7/2016, and DIP
STE1869/2-1 GE625/17-1. YB acknowledges ISF grant
1059/14. This work is supported in part by the facilities and staff of the Yale Center for Research Computing.

\section*{\nmr{Data Availability}}

\nmr{The data underlying this article will be shared on reasonable request to the corresponding author.}

\bibliographystyle{mn2e} 

\label{lastpage} 
 
\end{document}